\documentclass{elsart}
\usepackage[dvips]{epsfig}  
\usepackage{lineno}
\usepackage{graphicx}
\usepackage{xspace}         

\begin{document}
\newcounter{foot}
\newcounter{notes}
\newcommand{\cher}{Cherenkov \xspace}
\newcommand{\veff}{V_{\rm eff}}
\newcommand{\aeff}{A_{\rm eff}}
\newcommand{\aul}{$\overline{\mu}$}
\newcommand{\NN}{Neural Net}
\newcommand{\Q}{L_{\rm s}}
\newcommand{\drhoxy}{\Delta \rho_{\rm xy}}
\newcommand{\cosmu}{\cos \theta_{\mu}}
\newcommand{\rhoxy}{\Delta \rho_{\rm xy}}
\newcommand{\lvertex}{L_{\rm vertex}}
\newcommand{\lenergy}{L_{\rm energy}}
\newcommand{\zreco}{z_{\rm reco}}
\newcommand{\ereco}{E_{\rm reco}}
\newcommand{\rhoreco}{\rho_{\rm reco}}
\def\d{{\rm d}}

\newcommand{\stdunit}{\rm GeV ~cm^{-2}~ s^{-1} ~sr^{-1}}
\newcommand{\nbg}{$0.96^{+0.70}_{-0.43}$ \xspace}
\newcommand{\nmu}{$0.90^{+0.69}_{-0.43}$ \xspace}
\newcommand{\nuatmo}{$0.06^{+0.09}_{-0.04}$ \xspace}
\begin{frontmatter}
\title{Search for neutrino-induced cascades with AMANDA}
%
%
\vspace*{-0.1cm}
\author{
M.~Ackermann$^{4}$}
\author{J.~Ahrens$^{11}$} \author{ 
H.~Albrecht$^{4}$} \author{
X.~Bai$^{1}$} \author{ 
R.~Bay$^{9}$} \author{
M.~Bartelt$^{2}$} \author{
S.W.~Barwick$^{10}$} \author{ 
T.~Becka$^{11}$} \author{ 
K.H.~Becker$^{2}$} \author{
J.K.~Becker$^{2}$
% ...
}
\vspace*{-0.2cm}

\address{$^1$Bartol Research Institute, University of Delaware, Newark, DE 19716}
\address{$^2$Department of Physics, Bergische Universit\"at Wuppertal, D-42097 Wuppertal, Germany}
\address{$^3$Universit\'e Libre de Bruxelles, Science Faculty CP230, Boulevard du Triomphe, B-1050 Brussels, Belgium}
\address{$^4$DESY-Zeuthen, D-15735, Zeuthen, Germany}
\address{$^5$Blackett Laboratory, Imperial College, London SW7 2BW, UK}
\address{$^6$Dept. of Technology, Kalmar University, S-39182 Kalmar, Sweden}
\address{$^7$Lawrence Berkeley National Laboratory, Berkeley, CA 94720, USA}
\address{$^8$Dept. of Physics, Pennsylvania State University, University Park, PA 16802, USA}
\address{$^9$Dept. of Physics, University of California, Berkeley, CA 94720, USA}
\address{$^{10}$Dept. of Physics and Astronomy, University of California, Irvine, CA 92697, USA}
\address{$^{11}$Institute of Physics, University of Mainz, Staudinger Weg 7, D-55099 Mainz, Germany}
\address{$^{12}$Dept. of Physics, University of Maryland, College Park, MD 20742, USA}
\address{$^{13}$University of Mons-Hainaut, 7000 Mons, Belgium}
\address{$^{14}$Departamento de F\'{\i}sica, Universidad Sim\'on Bol\'{\i}var, Caracas, 1080, Venezuela}
\address{$^{15}$Dept. of Physics, University of Wisconsin, Madison, WI 53706, USA}
\address{$^{16}$Physics Dept., University of Wisconsin, River Falls, WI 54022, USA}
\address{$^{17}$Division of High Energy Physics, Uppsala University, S-75121 Uppsala, Sweden}
\address{$^{18}$Dept. of Physics, Stockholm University, SE-10691 Stockholm, Sweden}
\address{$^{19}$Vrije Universiteit Brussel, Dienst ELEM, B-1050 Brussels, Belgium}

\vspace*{-0.4cm}
\begin{abstract}
We report on a search for electro-magnetic and/or 
hadronic showers (cascades) induced by  high energy 
neutrinos in the data collected with the \mbox{AMANDA II} detector during 
the year 2000. The observed event rates are consistent with the expectations 
for atmospheric neutrinos and muons.
We place upper limits on a diffuse flux of extraterrestrial 
electron, tau and muon neutrinos. A flux of neutrinos with a spectrum 
$\Phi \propto E^{-2}$ which consists of an equal mix of all flavors,  
is limited to $E^2\Phi(E)=8.6 \times 10^{-7} \stdunit$ at a 90\% confidence 
level for a neutrino energy range 50~TeV to 5~PeV.  We present bounds for specific
 extraterrestrial neutrino flux predictions. Several of these models are ruled 
out. 
\end{abstract}
\end{frontmatter}

\section{Introduction}

The existence of high-energy extraterrestrial neutrinos is suggested 
by the observation of high-energy cosmic rays and gamma rays. 
Observation of neutrinos could shed light on the production and acceleration 
mechanisms of cosmic-rays, which for energies above the ``knee'' ($10^{15} \, 
\mathrm{eV}$) remain not understood.
Cosmic rays are thought to be 
accelerated at the shock fronts of galactic objects 
like supernova remnants, micro-quasars, and in extragalactic 
sources such as the cores and jets of active galactic nuclei (AGN) \cite{halzen02}. 
High energy protons accelerated in these objects may collide with the 
 gas and
radiation surrounding the acceleration region, or with matter or radiation 
 between the source and the Earth.
Charged pions, produced in the interaction,  decay into 
highly energetic muon neutrinos and muons which further decay into electron 
neutrinos.  
Fermi acceleration of charged particles in magnetic shocks naturally leads 
to power-law 
spectra, $E^{-\alpha}$, where $\alpha$ is typically close to 2. 
Hence, the spectrum of  astrophysical neutrinos is harder than the spectrum
of atmospheric neutrinos ($\sim E^{-3.7}$) potentially allowing to distinguish 
the origin of the flux (see for example \cite{lm}).

For a generic astrophysical neutrino source, one expects a ratio of neutrino
fluxes $\Phi_{\nu_e}:\Phi_{\nu_\mu}:\Phi_{\nu_\tau}\approx 1:2:0$. Due to neutrino vacuum oscillations this ratio changes to 
$\Phi_{\nu_e}:\Phi_{\nu_\mu}:\Phi_{\nu_\tau} \approx1:1:1$ by the time the 
neutrinos reach the Earth. 
Recently a search with the AMANDA detector was reported \cite{gary},
 resulting in the most restrictive upper 
limit on the diffuse flux of muon neutrinos (in the energy range  6 to 
1000 TeV). Clearly, a high sensitivity to 
neutrinos of all neutrino flavors is 
desirable. The present paper reports on a search for a diffuse flux of 
neutrinos of all flavors performed using neutrino-induced cascades in AMANDA.

\section{The AMANDA Detector}

AMANDA-II \cite{nature} is a Cherenkov detector consisting of 677 
photomultiplier tubes (PMTs) arranged on 19 strings. It is frozen into the 
Antarctic polar ice cap 
at a depth ranging mainly from 1500~ to 2000~m. AMANDA detects high-energy 
neutrinos by observation of the \cher light from charged particles 
produced in neutrino interactions. The detector was triggered when the number 
of PMTs with signal (hits) reaches 24 within a time-window of 2.5~$\mu$s.

The standard signatures are 
neutrino-induced muons from charged current (CC) $\nu_\mu$ interactions.
The long range of high energy muons, which leads to large detectable signal 
event rates and good angular resolution results in restrictive bounds on
neutrino point-sources \cite{psp}. 

Other signatures are hadronic and/or electro-magnetic cascades generated by 
CC interaction of $\nu_e$ and $\nu_\tau$.
Additional cascade events from all neutrino flavors are
obtained from neutral current interactions.
Good energy resolution, combined with low background
from atmospheric neutrinos makes the study of cascades  a feasible
method to search for extraterrestrial high energy neutrinos.

\section{Update on Cascade Search with AMANDA-B10}

Before the completion of AMANDA-II, the detector was operated in a smaller 
configuration. The results for the search of neutrino induced cascades in 130.1
effective days of the 10-string AMANDA-B10 detector during 1997 have been
reported before \cite{b10}. The same analysis has been
applied to 221.1 effective days of experimental data collected during
1999. The AMANDA detector in 1999 had three more strings than in 1997,
yet data from these strings were not used in this analysis, so that
the detector configuration used in the 1999 neutrino induced cascade
search is very similar to that of 1997.

Signal simulation for the analysis of 1999 data
was improved to the standards reported in this letter. 
No events were found in the
1999 experimental data after all selection criteria had been
applied. We will present results supposing a
background of zero events.

Using the procedure explained in this letter we obtain an upper limit on the 
number of signal events of $\mu_{90\%}$=2.75 at a 90\% confidence level, from 
which 
we calculate the limit on the flux of all neutrino flavors. Assuming a flux 
$\Phi \propto E^{-2}$ consisting of an equal mix of all flavors, one obtains an 
 upper limit $ \Phi_{90\%} = 8.9 \times 10^{-6}\;\mathrm{GeV\,cm^{-2}\,s^{-1}\,sr^{-1}}$.
 In the calculation
of this limit we included a systematic uncertainty on the signal detection efficiency 
of $\pm$32\%. About 90\% of the simulated signal events for this limit have 
energies
between 5 and 300 TeV, while 5\% have lower and 5\% have higher energy.
Differences between this result and the one obtained with 1997
experimental data \cite{b10} are due to the larger live-time in 1999 and improved
simulation.

\section{Data Selection and Analysis for AMANDA-II}

The data set of the first year of AMANDA-II operation comprises 
$1.2\times 10^{9}$ triggered events collected over 238 
days between February and November, 2000, with 197 days live-time after 
correcting for detector dead-time. 

The background of atmospheric muons was simulated
with the air-shower simulation program CORSIKA~(v5.7)~\cite{corsika} using the 
average winter air density at the South Pole and the QGSJET hadronic 
interaction model \cite{qgsjet}. The cosmic ray composition was taken from 
\cite{ws}. All muons were propagated through the ice 
using the muon propagation program MMC (v1.0.5)~\cite{mmc}.
The simulation of the detector response includes the propagation of \cher photons
through the ice as well as the response of the PMTs and the surface electronics.

Besides generating unbiased background events, the 
simulation chain was optimized to the higher energy threshold of this analysis.
By demanding that atmospheric muons passing through the detector radiate a secondary
with an energy of more than 3~TeV, the simulation speed is increased significantly. 
A sample equivalent to 920 days of atmospheric muon 
data was generated with the optimized simulation chain.

The simulation of $\nu_e, \nu_\mu$ and $\nu_\tau$ events was done 
using the signal generation program ANIS (v1.0)~\cite{anis}.
The simulation includes CC and 
neutral current (NC)interactions as well as $W^-$ production in the $\overline{\nu}_e e^-$ 
channel  near 6.3~PeV (Glashow resonance). All relevant neutrino 
propagation effects inside the Earth, such as neutrino absorption or 
$\nu_\tau$ regeneration are included in the simulation.

The data were reconstructed with methods described in Ref.\ \cite{b10}. 
Using the time information of all hits, a likelihood 
fit results in a 
 vertex resolution for cascade-like events of about 5~m in the 
 transverse coordinates (x,y) and slightly better in the depth 
coordinate (z).
The reconstructed vertex position combined with a model for the
 energy dependent hit-pattern of cascades allows the
 reconstruction of the energy of the cascade using a likelihood method.
The obtained energy 
resolution in $\log_{10}E$ lies between 0.1 and 0.2. The performance of the
reconstruction methods have been verified using {\it in situ} light sources.
 
\begin{table}[htbp]
\begin{center}
\begin{tabular}{ c c c   c   c c  c  }
\hline
\hline
\#&  cut variable &exp. & & & MC & \\ 
& & & & atm. $\mu$  & atm. $\nu_e$  & $E^{-2}$ $\nu_e$   \\
\hline
1&$N_{early}/N_{hit}<0.05$ & 0.058 & & 0.033 & 0.94  &0.63 \\
2&$N_{dir}>8$ & 0.030 & & 0.016 & 0.89 & 0.57 \\
3 &$L_{\rm vertex}<7.1$ & 0.0027 & & 0.0012 & 0.39  & 0.35 \\
4 &$\lenergy vs. E_{\rm reco}$ & 0.0018 & & 0.00077 & 0.35  & 0.26 \\
5 &$-60<\zreco<200$ & 0.0010 & & 5.9$\cdot 10^{-4}$ & 0.28 & 0.18 \\
6 &$\rhoreco vs. E_{\rm reco}$ & 8.6$\cdot10^{-4}$ & & 5.1$\cdot 10^{-4}$ & 0.26  & 0.15 \\
7 &$\Q>0.94$ & 9.7$\cdot 10^{-6}$ & & 4.8$\cdot 10^{-6}$ & 0.040  & 0.091 \\
8 &$E_{\rm reco}>50~{\rm TeV}$ & 8$\cdot 10^{-10}$ & & 7$\cdot 10^{-10}$
 & 2.8$\cdot10^{-5}$ & 0.029  \\
\hline
\hline
\end{tabular}
\caption[Fraction of triggered events passing the cuts of this analysis]{
Cumulative fraction of triggered events passing the cuts of this analysis. Values are 
given for experimental data, atmospheric muon background Monte Carlo (MC) simulation, atmospheric 
$\nu_e$ simulation and a $\nu_e$ signal simulation with an energy  
spectrum $\Phi \propto E^{-2}$.
The flavor $\nu_e$ was chosen to illustrate the filter 
efficiencies, since interactions of $\nu_e$ always lead to cascade-like 
events. 
}
\label{tab:level1}
\label{tab:level2}
\end{center}
\end{table}

Eight cuts were used to reduce the background from atmospheric muons by a 
factor $\sim 10^{9}$. The different cuts are explained below.
The cumulative fraction of events that passed the filter steps 
are summarized
in Table \ref{tab:level1}. 

Since the energy spectrum of the background  is falling steeply one obtains 
large systematic uncertainties from threshold effects in this analysis. For 
example, 
an uncertainty of $\pm$30\% in the photon detection efficiency translates to 
up to a 
factor $2^{\pm 1}$ uncertainty in rate. Such effects can explain the 
discrepancies of Table \ref{tab:level1} in passing efficiencies between 
atmospheric 
muon background simulation and experimental data. However, as will be shown later, the 
threshold effects are smaller for harder signal-like spectra.

At the lowest filter levels (cuts 1 and 2), variables based on
 a rough {\it first-guess} vertex position reconstruction are used to 
 reduce the number of background events by about a factor of 30.
It is useful to define the time residual of a hit as the time delay of the 
hit time relative to the 
time expected from unscattered photons.
The number of hits with a negative time residual, $N_{\rm early}$, divided by the 
number of all hits, $N_{\rm hits}$, in an event should be small. This first cut 
criterion is 
effective since  early hits are not consistent with the expectation from 
cascades, while they are expected from long muon tracks.
Cut 2 enforces that the number of so called direct 
hits, $N_{dir}$ (photons having a time residual between 0 and 200~ns), is 
large.

Cut 3 is a requirement on the
reduced likelihood parameter resulting from the standard vertex fit, $\lvertex<7.1$ (see also \cite{b10}). Note, that the likelihood parameter is, in analogy to a reduced $\chi^2$, 
defined such that  smaller values indicate  a better fit result, hence a 
more signal-like event. In a similar manner, the resulting likelihood value from the energy fit, $\lenergy$, is used as a selection criterion (cut 4). 
However, since  the average value of $\lenergy$ has an energy dependence, 
the cut value is  a function of the reconstructed energy, $E_{\rm reco}$.
Cut 5 on the reconstructed $z$ coordinate, 
$z_{\rm reco}$, was introduced
to remove events which are reconstructed outside AMANDA and in regions where 
the simulation of the ice properties for photon propagation is insufficient. 
While the upper boundary coincides roughly 
with the detector boundary, the lower value is about 100~m above the
geometrical border of the detector. Restricting $z_{\rm reco}$ improves
significantly the description of the remaining experimental data 
(for example the reconstructed energy spectrum) \cite{mk_phd}. 
\begin{figure}[htb]
\begin{center}
\includegraphics[width=0.4\textwidth]{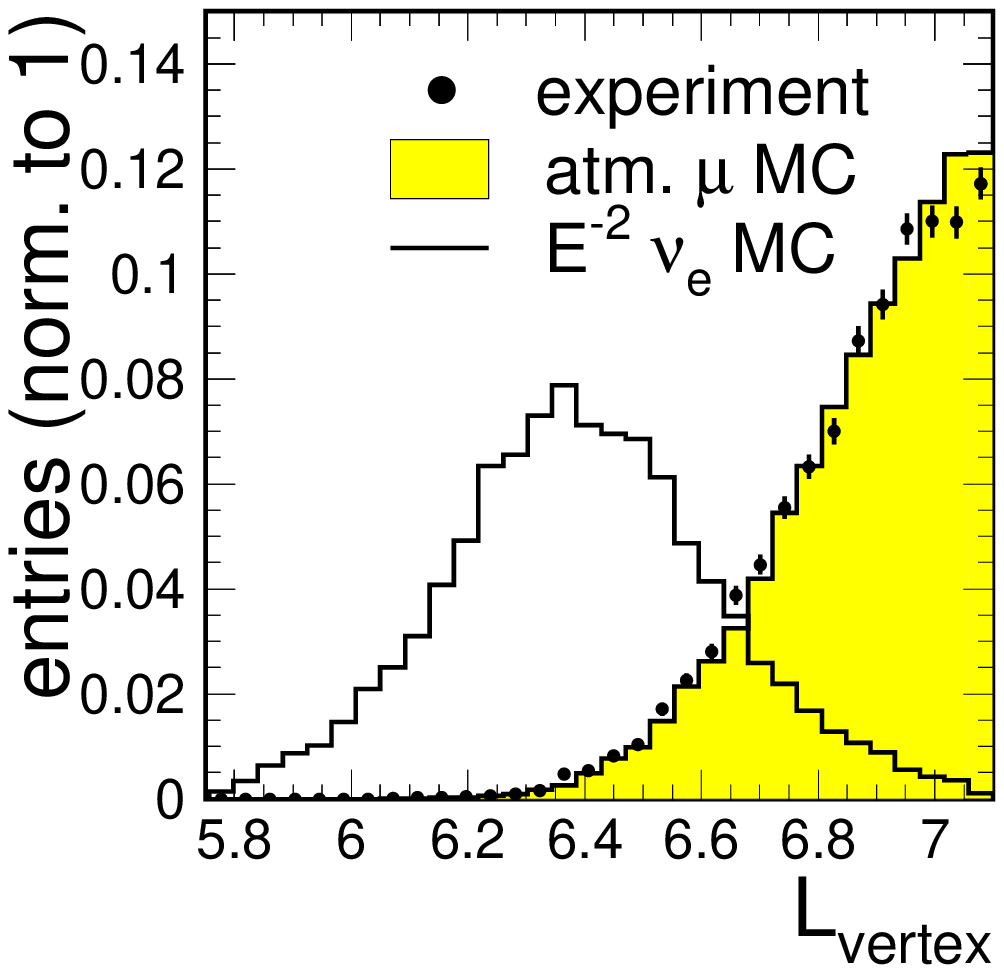}  
\includegraphics[width=0.4\textwidth]{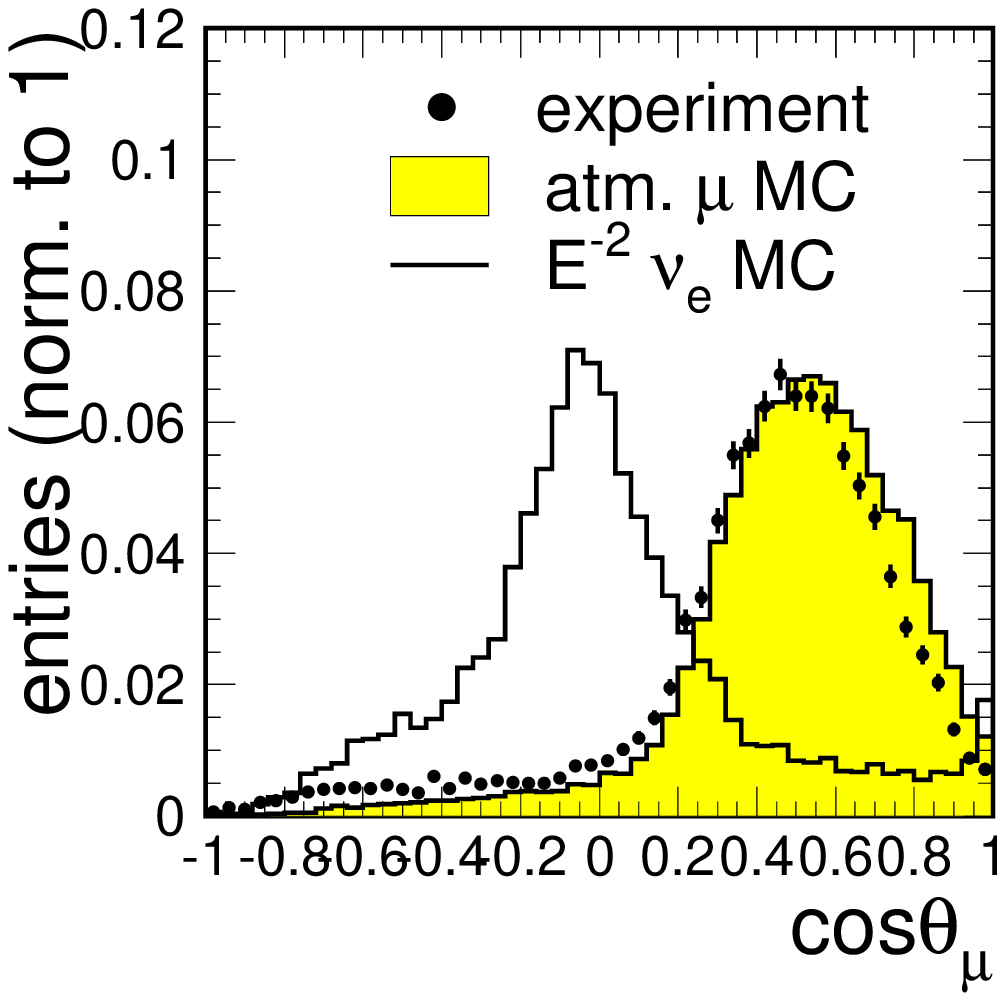}  
\includegraphics[width=0.4\textwidth]{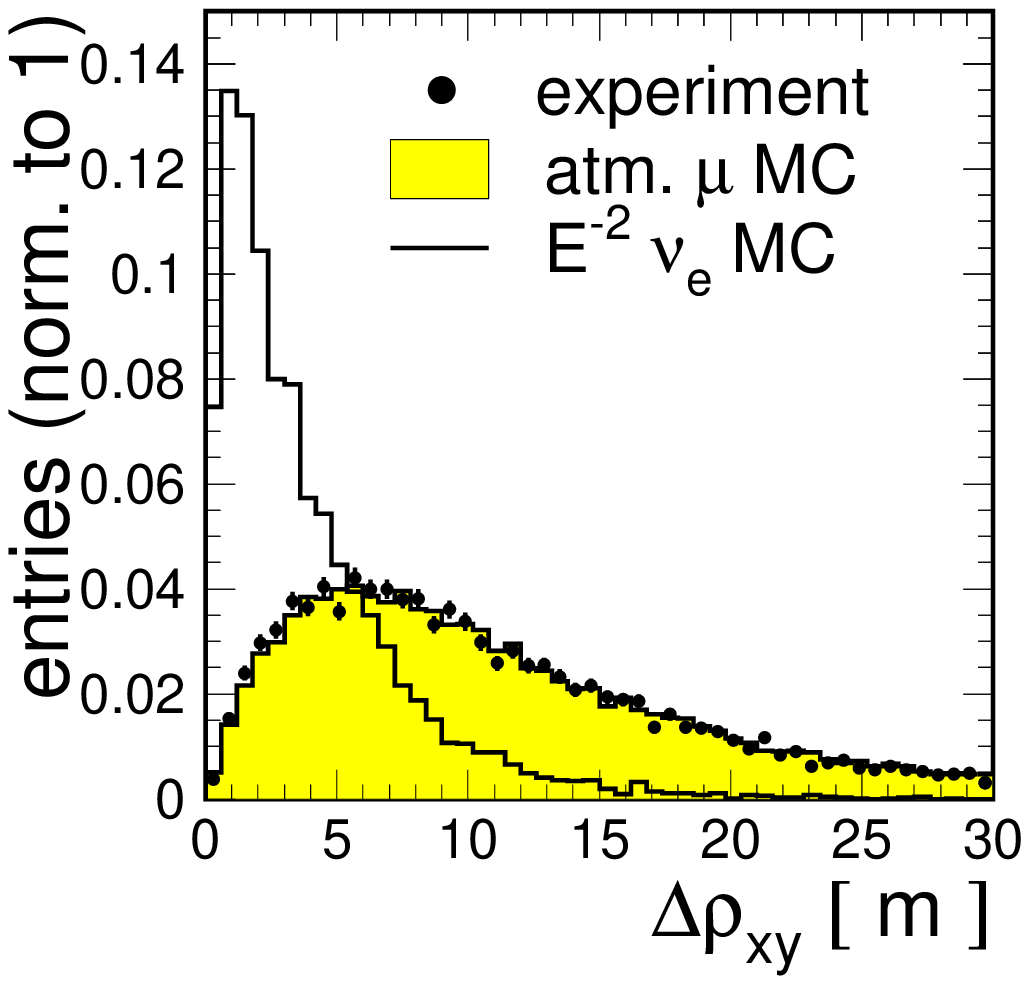}  
\includegraphics[width=0.4\textwidth]{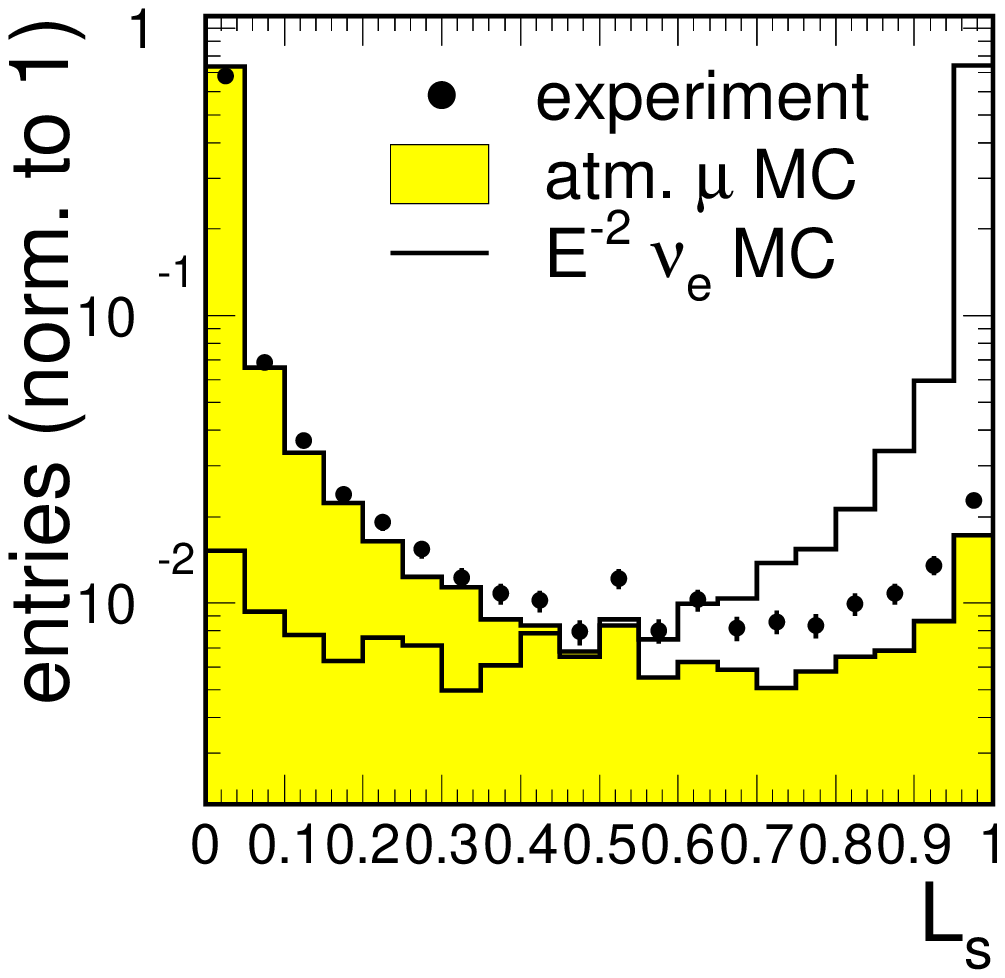}  
\end{center}
\caption{Normalized distribution of the three input variables 
$\lvertex$, $\cosmu$ and  $\rhoxy$ as well as the resulting likelihood 
variable $\Q$.  Shown are experimental data as well as
atmospheric muon and signal MC simulations after cut 6.}
\label{like}
\end{figure} 
Only events reconstructed with a radial distance to the detector z-axis, 
$\rhoreco < 100$~m, are accepted (cut 6), unless their reconstructed energies 
lie above 10~TeV. 
For each decade in energy above 10~TeV one 
allows the maximal radial distance to grow by 75~m. 
This reflects the fact that the  
cascade radius\footnote{We define the cascade event radius as the direction 
averaged distance from the vertex at which the average number of registered 
photon-electrons is equal to 1.}, 
increases as a function of 
energy, while the expected
amount of background decreases.

Three discriminating variables are used to form the final quality
parameter $\Q$:

\begin{itemize}
\item[1.]{The value of the 
reduced likelihood parameter resulting from the vertex fit, $\lvertex$. 
Note that this variable has  been used previously in cut 3.
}

\item[2.]{ The difference in the radial distance of the vertex position 
reconstructed with two different hit samples, $\rhoxy$. While the first reconstruction 
is the 
regular vertex reconstruction using all hits, the second reconstruction 
uses only those 
hits outside a 60~m sphere around the vertex 
position resulting from  the first reconstruction. Since the close-by hits 
typically 
contribute most to the likelihood function, their omission allows to test 
the stability of the reconstruction result.  
If the underlying event is a 
neutrino-induced cascade, the second 
reconstruction results in a  
vertex position close to that of the first reconstruction.
In case of a misidentified muon event, removing hits located close to the 
vertex typically results in a significantly different reconstructed position.  
}

\item[3.]{ The cosine of the angle of incidence $\cosmu$ as reconstructed with a muon-track fit.
The muon-track fit assumes for the underlying likelihood parametrization 
that the hit pattern  originates from a long range muon track. 
The fit allows to reconstruct correctly a large fraction of the atmospheric 
muons. }

\end{itemize}
The final quality parameter  is defined as a likelihood ratio: 
\begin{equation}
\Q =\frac{\prod_i p^{s}_i(x_i)}
 {\prod_i p^{s}(x_i) +\prod_i p^{b}(x_i)},
\end{equation}  
where $i$ runs over the
 three variables. $p^h$ ($h=s$ for signal and $h=b$ for background) 
are probability density functions defined as $p^h(x_i) =
f_i^{h}(x_i)/(f_i^{s}(x_i) + f_i^{b}(x_i))$. $f^h(x_i)$ correspond to the 
probability density functions of the individual variables $x_i$ for
background
due to atmospheric muons  and signal consisting of a flux of 
$\nu_e$ 
with a spectral slope $\Phi(E_\nu) \propto E_\nu^{-2}$. They 
are obtained from  simulations.

The distributions of the individual variables as well as of the likelihood 
ratio $\Q$  are shown in 
Fig.\ \ref{like} for experimental data, atmospheric muon background and signal 
simulations. The experimental distributions of $\rhoxy$ and $\lvertex$ 
approximately  agree with those from the simulation while the distribution 
of 
$\cosmu$ shows some larger deviations. The deviation reflects an 
simplified description of the photon propagation through the dust layers in 
the ice \cite{mk_phd}.    
The experimental $\Q$ distribution is not perfectly described by 
the atmospheric muon simulation, which is mainly related to the mis-match
in the $\cosmu$ distribution.
The related uncertainties in the cut efficiencies are included
in the final results.


At this stage of the event selection  one is left with events due to
atmospheric muons, which happen to radiate (mostly through bremsstrahlung) a 
large fraction of their energy into a single electro-magnetic cascade.
The reconstructed energy corresponds to that of the 
most energetic secondary-particle cascade produced in the near vicinity of the detector. 
To optimize the sensitivity of the analysis to an astrophysical 
flux of neutrinos, a further cut on the reconstructed energy,  $E_{\rm reco}$, 
was introduced. 

\begin{figure}[htb]
\begin{center}
\includegraphics[width=0.5\textwidth]{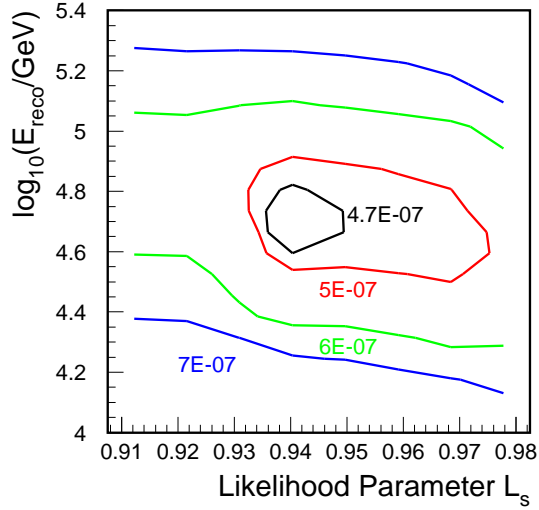} 
\end{center}
\caption{Optimization of final cuts.
The sensitivity for the diffuse flux of $\nu_e$ is shown as a function of
cuts on $\ereco$ and $\Q$. The coefficient next to the contour lines correspond
to the average upper limit in units of $(E/\rm GeV)^{-2} \cdot {\rm GeV^{-1}~ s^{-1} sr^{-1} cm^{-2}}$. }
\label{cont}
\end{figure} 

The sensitivity is defined as the average upper limit on the neutrino flux 
obtained from a large 
number of identical experiments in the absence of signal \cite{mrp,fc98}. 
The sensitivity was calculated for a flux of $\nu_e$ with spectrum 
$\propto E^{-2}$. A flux of $\nu_e$ was used 
for optimization, since $\nu_e$-induced events always have  cascade-like 
signatures. 
The sensitivity is shown 
in Fig.\ \ref{cont} as a function of the $E_{\rm reco}$ and $\Q$ cut. 
$\Q>0.94$ and $E_{\rm reco}>50$~TeV  were chosen in this two dimensional 
optimization 
procedure such that the average upper limit is lowest.   
With these cuts the expected sensitivity for an ~$E^{-2}$ spectrum of 
electron neutrinos is
 $4.6\times10^{-7} (E/\rm GeV)^{-2} \cdot {\rm GeV^{-1}~ s^{-1} sr^{-1} cm^{-2}}$. 

\begin{figure}[htb]
\begin{center}
\includegraphics[width=0.5\textwidth]{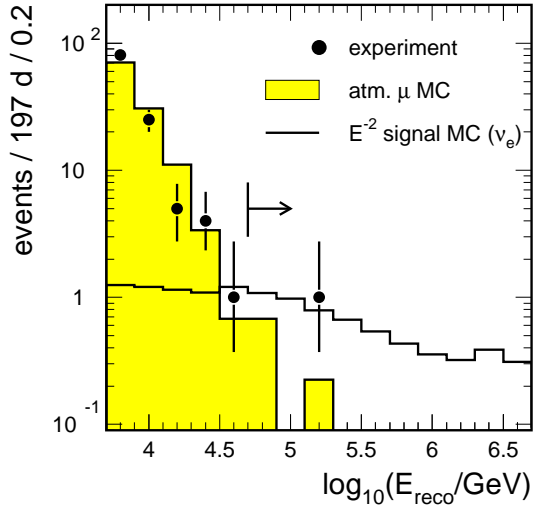} 
\end{center}
\caption{Distributions
  of reconstructed energies after all but the final energy cut. Shown are
experimental data, atmospheric muon simulation and a hypothetical
flux of astrophysical neutrinos. The final energy cut is indicated by the
line with the arrow.}
\label{finalspec}
\end{figure} 

The energy spectra of experimental data as well as signal and background 
simulations after all but the final energy cut are shown in 
Fig.\ \ref{finalspec}. Note that the energy spectrum begins at 5 TeV, since 
this is the lowest energy for which the optimized background simulation is 
applicable. The number of events due to simulated atmospheric muons was 
normalized to that observed in the experiment.

One experimental event passes all cuts, while \nbg events are expected 
from atmospheric muons and a small contribution from 
atmospheric neutrinos. 

The spectrum  as obtained from simulation of atmospheric muons passing cut 7 
was normalized to the number of experimental events resulting in 
an expectation of \nmu events due to atmospheric muons. The three main 
sources to the error are given by limited statistics of simulated 
atmospheric muon events (the error of $^{+0.65}_{-0.36}$ was 
determined using the 
Feldman-Cousins method \cite{fc98}), uncertainties in the cut efficiency 
($\pm20\%$ obtained from variation of the cuts) and limited 
knowledge of the ice properties ($\pm 12\%$ obtained from variation of the ice properties in the simulation).  The total error was obtained by adding the 
individual errors in quadrature.

The predicted event number from 
atmospheric neutrinos simulated according the flux of Lipari 
\cite{lipari} is \nuatmo, where the uncertainties are mainly due to 
uncertainties in  ice properties (error of $\pm0.03$ obtained from variation of the ice properties in simulation), and in detection efficiencies of  
\cher photons ($^{+0.08}_{-0.02}$ obtained from variation of the 
photon detection sensitivity in the simulation). 
The theoretical uncertainties in the flux of 
atmospheric neutrinos was estimated to be about 25\%~\cite{gaisser} and is small when compared with the other uncertainties. 
Again, the total error was obtained by adding the 
individual errors in quadrature.

The uncertainty in the 
detection efficiency of neutrino events from an astrophysical
flux with a spectral index $\alpha\leq2$ are estimated to be not 
larger than 25~\%. Because of the flatter energy spectrum, the uncertainties
related to the energy threshold (such as the photon detection efficiency)
result in smaller uncertainties in rate when compared 
to the uncertainties found for atmospheric neutrino events. The main 
sources of error are again uncertainties in the simulation of the ice 
properties ($\pm15\%$) and the detection efficiencies of the \cher photons ($\pm20\%$).

\begin{figure}[htp]
\begin{center}
\begin{minipage}{0.3\textwidth}
\epsfig{file=e_s3.epsi,width=1\textwidth,clip=1}
\end{minipage}
\hfill
\begin{minipage}{0.43\textwidth}
\epsfig{file=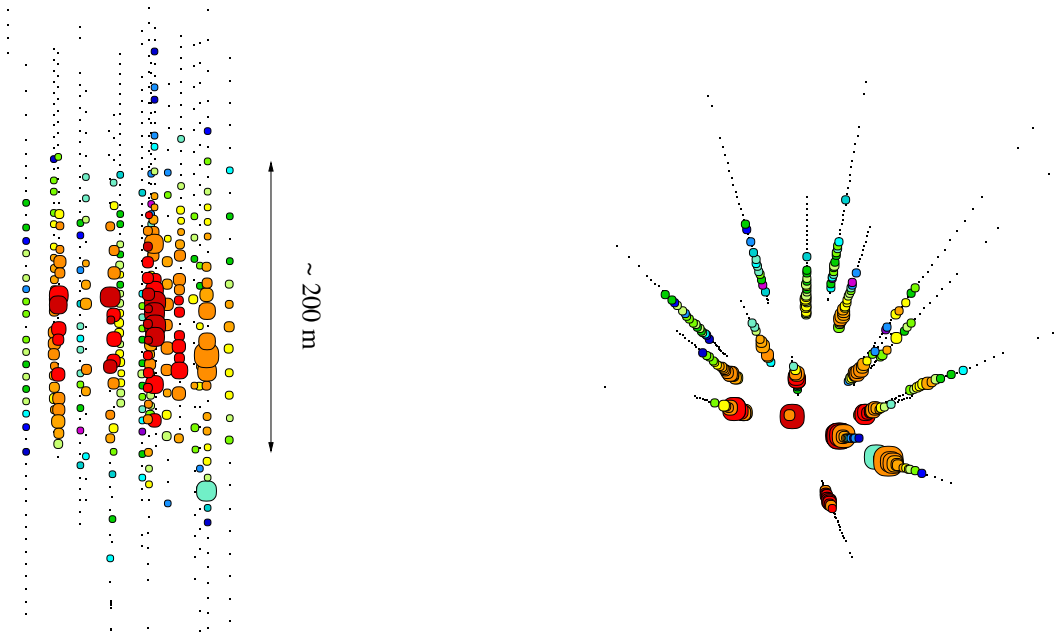,width=1\textwidth,clip=1}
\end{minipage}
\caption[The remaining event]{The experimental event which has passed all selection criteria is displayed from the side (left) and from above (right).  Points represent PMTs, and shaded circles represent hit PMTs 
(early hits have  darker shading, late hits have lighter shading). Larger 
circles represent larger registered amplitudes. The light pattern  has the 
sphericity and time profile expected from a neutrino induced cascade. The arrow  indicates the length scale.
}
\label{event}
\end{center}
\end{figure}

The experimental event which passed all selection criteria is shown in 
Fig.\ \ref{event}.

The sensitivity of the detector to neutrinos can be
characterized by its effective volume, $\veff$, or area, $\aeff$, remaining 
after all cuts are applied.  $\veff$ represents the volume, 
in which neutrino interactions are observed with full efficiency while 
$\aeff$ represents the area with which a neutrino flux can be observed with 
full efficiency. While the concept of $\veff$ is more intuitive because it
relates to the geometrical size of the detector, the concept of $\aeff$ is
more convenient for calculations of neutrino rates (see Eq. \ref{eq:events} in Sect. \ref{sec:results}).

\begin{figure}[htb]
\begin{center}
\includegraphics[width=0.8\textwidth]{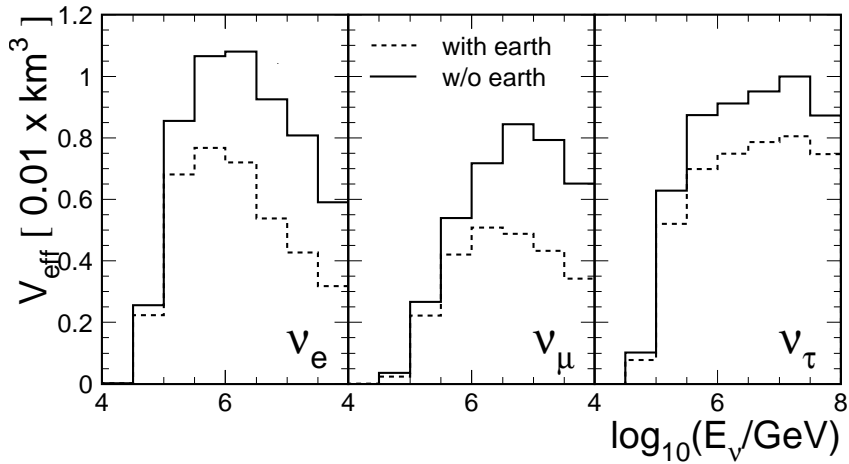} 
\end{center}

\caption{Effective volume for $\nu_e,\nu_\mu$ and
$\nu_\tau$ interactions as a function of the neutrino energy. The effective volume is shown without including Earth propagation effects (full line) and with  Earth propagation effects (dashed line). }
\label{effvol}
\end{figure}

\begin{figure}[htb]
\begin{center}
\includegraphics[width=0.8\textwidth]{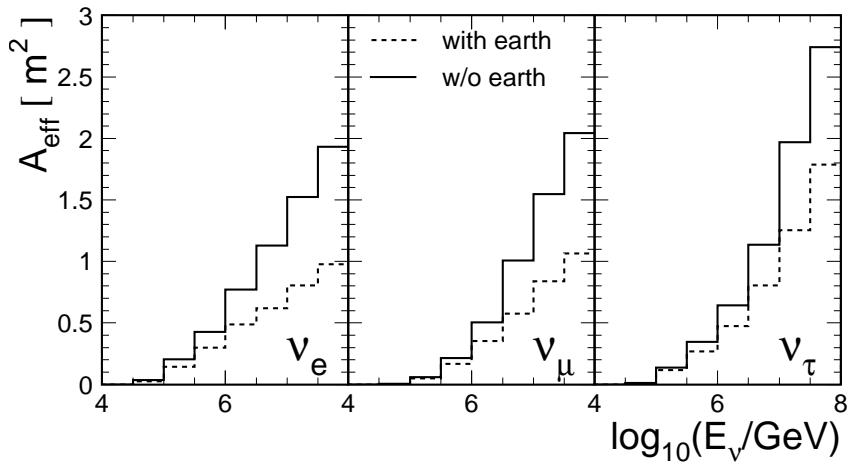} 
\end{center}

\caption{Effective area for $\nu_e,\nu_\mu$ and
$\nu_\tau$ interactions as a function of the neutrino energy. The effective area is shown without including Earth propagation effects (full line) and with  
Earth propagation effects (dashed line). }
\label{effarea}
\end{figure}

Figure \ref{effvol} shows $\veff$  as obtained from simulation 
for all 
three neutrino flavors as a function of the neutrino energy. 
The effective volume has been averaged over all neutrino arrival directions.
As can be seen,  
$\veff$ rises for energies above the threshold energy of 50 TeV. 
Above PeV-energies $\veff$ decreases for $\nu_e$ and $\nu_\mu$, an effect 
related to both reduced filter efficiencies and neutrino absorption effects. 
In the case of $\nu_\tau$, the volume saturates
because of regeneration effects: $\nu_\tau\rightarrow\tau\rightarrow\nu_\tau$
and because of the event $\nu_\tau$ event topology: 
there is an increase in detection probability for
CC $\nu_\tau$ interactions (with energies above $\sim 10^7$~GeV) 
because the cascade from the hadronic vertex and 
the cascade arising from the subsequent tau decay are separated far enough 
in space to be detected independently.

Fig.\ \ref{effarea} shows $\aeff$  as obtained from simulation 
for all three neutrino flavors as a function of the neutrino energy. 
Note that $\aeff$ is small because of the small 
neutrino interaction probability, which is included in the calculation of 
$\aeff$ (but not in $\veff$).

The detector sensitivity varies only weakly as a function of
the neutrino incidence angles. However, because of neutrino propagation 
effects effective area and volume are suppressed for neutrinos coming from 
positive declinations.

The effect of the 
resonant increase of the cross-section for $\bar{\nu}_e$ at the Glashow 
resonance is not included in  
$\aeff$ shown in Fig.\ \ref{effarea}. For energies between 
$10^{6.7}$ and $10^{6.9}$~GeV
the average effective area including Earth propagation effects is 
$\overline{A_{\rm eff}^{\overline{\nu}_e}}=8.4 ~{\rm m}^2$.

\section{Results}
\label{sec:results}
\begin{table}[htbp]
\begin{center}
\begin{tabular}{ l c c c c c }
\hline
\hline
Model & $\nu_e$ & $\nu_\mu$ & $\nu_\tau$ &  $\nu_e+\nu_\mu+\nu_\tau$
&$\frac{\mu_{90\%}}{N_{\rm model}}$\\ 
\hline
$10^{-6}\times E^{-2}$   & 2.08 & 0.811 & 1.28 & 4.18 & 0.86 \\       
 SDSS  \cite{stecker1}                  & 4.20 & 1.91 & 2.77 & 8.88 & 0.40 \\
 SS Quasar \cite{stecker2}              & 8.21 & 3.57 & 5.30 & 17.08 & 0.21 \\
 SP u   \cite{sp}                 & 33.0 & 13.0 & 20.5 & 66.6 & 0.054 \\
 SP l  \cite{sp}                  & 6.41 & 2.34 & 3.98 & 12.7 & 0.28 \\
  P $pp+p\gamma$  \cite{P96}        & 5.27 & 1.57 & 2.86 & 9.70 & 0.37 \\
 P $p\gamma$     \cite{P96}        & 0.84 & 0.40 & 0.56 & 1.80 & 1.99 \\
MPR   \cite{mpr}                  & 0.38 & 0.18 & 0.25 & 0.81 & 4.41 \\
\hline
\hline
\end{tabular}
\caption[Event rates and model rejection factors for various models]{Event
  rates and model rejection factors (MRF) for models of astrophysical neutrino
  sources. The assumed upper limit on the number of signal events with all 
uncertainties incorporated is   $\mu_{90\%}=3.61$}
\label{tab:model_limits}
\end{center}
\end{table}

Since no excess events have been observed above the expected backgrounds, 
 upper limits on the flux
of astrophysical neutrinos are calculated. The uncertainties in both 
background expectation and signal efficiency, as discussed above, are 
included in the calculation of the upper limits.
We assume a mean  background of 0.96 with a Gaussian distributed 
relative error of 73\%, and an error on the signal 
detection efficiency of 25\%. 
For a 90\% confidence level an upper limit on the number of signal events,  $\mu_{90\%}=3.61$,
is obtained using the Cousins-Highland \cite{ch92} prescription implemented by Conrad {\it et al.} \cite{conrad}, with 
the unified Feldman-Cousins ordering \cite{fc98}. Without any
uncertainties the upper limit on the number of signal events would be 3.4. 

The effective area can be used to calculate the expected event numbers for any 
assumed flux of neutrinos of flavor $i$, $\Phi_i(E_\nu)$:
 
\begin{equation}
N_{\rm model} =4\times \pi\times T \sum_{i=\nu_e,\nu_\mu,\nu_\tau} \int \d E_\nu~\Phi_i(E_\nu) \aeff^i(E_\nu),
\label{eq:events}
\end{equation}
with $T$ being the live-time. 
If $N_{\rm model}$ is larger than $\mu_{90\%}$, the model is ruled out at 
90\% CL. Table \ref{tab:model_limits} summarizes the predicted event numbers 
for different models of hypothetical neutrino sources. Thereby,  the spectral 
forms of $\nu_\mu$ and $\nu_e$ are assumed to be the same (the validity of 
this  approximation is discussed in \cite{mk_phd}). Furthermore,  full mixing 
of neutrino flavors is assumed, hence 
$\Phi_{\nu_e}:\Phi_{\nu_\mu}:\Phi_{\nu_\tau}= 1:1:1$ as well as a ratio 
$\nu / \bar\nu=1$.

\begin{figure}
\begin{center}
\includegraphics[width=0.5\textwidth]{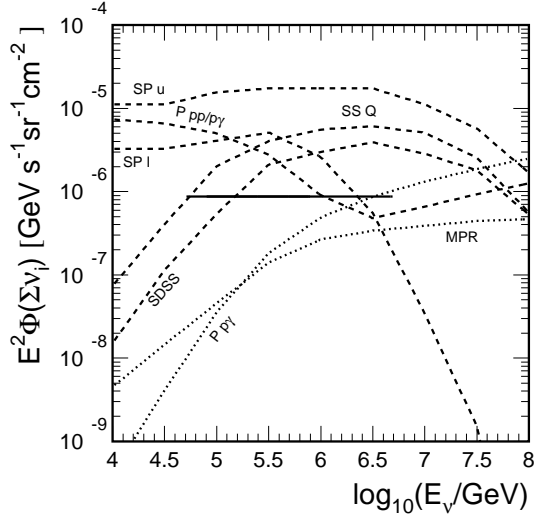} 
\end{center}
\caption{Flux
  predictions for models of astrophysical neutrinos sources. 
Models represented by dashed lines are excluded by the results of this work.
Models fluxes represented by dotted lines are consistent with the experimental 
data. The labels are
explained in the text. The solid line corresponds to the upper limit 
on a flux $\Phi\propto E^{-2}$.}
\label{limits}
\end{figure}

Electron neutrinos contribute about 50\% to the total event 
rate, tau neutrinos about 30\% and muon neutrinos about 20\%.   
For the sum of all neutrino flavors the various predicted fluxes 
are shown in Fig.\ \ref{limits}.

The models by Stecker {\it et al.} \cite{stecker1} labeled ``SDSS'' and its update \cite{stecker2} ``SS Q'', 
as well as the models by Szabo and Protheroe \cite{sp} ``SP~u'' and ``SP~l'' represent 
models for  neutrino production in the central region of Active Galactic 
Nuclei. As can be seen from Table \ref{tab:model_limits}, 
these models are ruled out with 
$\frac{\mu_{90\%}}{N_{\rm model}}\approx 0.05-0.4$. 
Further shown are models for neutrino production in AGN jets: 
a calculation by Protheroe \cite{P96}, which 
includes neutrino 
production through $p\gamma$ and $pp$ collisions 
(models ``P $pp+p\gamma$'' and ``P~$p\gamma$'') as well as an 
evaluation of the maximum flux 
due to a superposition of possible extragalactic sources by 
Mannheim, Protheroe and Rachen \cite{mpr} (model ``MPR'').  
The latter two models are currently not excluded. 

For a neutrino flux of all flavors with spectrum $\propto E^{-2}$ one obtains 
the limit:
\[ E^2 \Phi_{90\%}=8.6\times 10^{-7} ~\stdunit.\] 
For such a spectrum, about 90\% of the events detected have neutrino energies 
between 50~TeV and 5~PeV,with 
the remainder equally divided between the ranges above and 
below. The limit is shown in Fig.\ \ref{limits} as a solid line 
ranging from  50~TeV to 5~PeV.

\begin{figure}
\begin{center}
\includegraphics[width=0.5\textwidth]{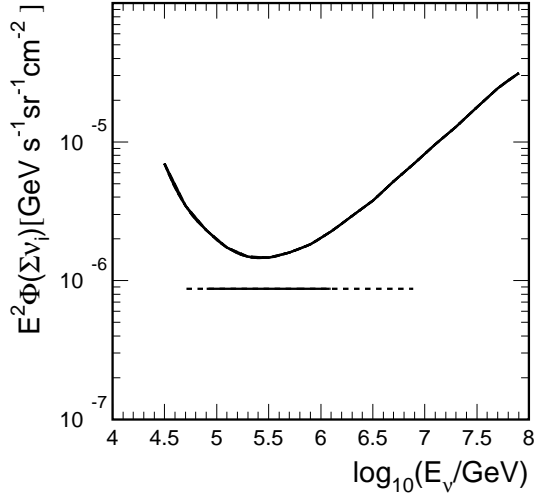} 
\end{center}
\caption{Illustration of the energy dependency. The curved solid line represents 
a superposition of the 
limiting fluxes of a series of power law models 
$\Phi \propto E^{-2}$ restricted to one decade in energy. 
The limit for a flux  $\Phi \propto E^{-2}$ 
without energy restriction is shown for comparison. 
The range of the dashed line represents 
the range of energies in which 90\% of the signal events are detected while
the embedded solid line represents the energy range 
in which 50\% of the signal events are detected, with 
the remainder equally divided between the ranges above and below.}
\label{limit2}
\end{figure}

To illustrate the energy dependent sensitivity of the present
analysis we restrict the energy range for integration of Eq.
(\ref{eq:events}) to one decade.
By assuming a 
benchmark flux $\Phi_{E_0}(E)= \Phi_0 \times (E/E_0)^{-2} \times 
\Theta(0.5-|\log(E/E_0)|)$ where $\Phi_0=1/({\rm GeV~cm^2~s~sr})$ represents 
the unit flux and $\Theta$ the Heaviside step-function (restricting the energy 
range to one decade), one 
obtains the number of events for a given central energy $E_0$: 
$N_{\rm event}(E_0)$. 
The limiting flux at the energy $E_0$ is then given by 
$\Phi_{90\%}(E_0)=\Phi_0\times\mu_{90\%}/N_{\rm event}(E_0)$. The 
superposition of the limiting fluxes as a function of the central energy  
is shown in Fig.\ \ref{limit2}. For a flux 
$\Phi\propto E^{-2}$ the analysis 
has its largest sensitivity around 300~TeV.

The mentioned strong increase in effective area at the energy of the Glashow 
resonance
allows setting of a limit on the differential flux of $\overline{\nu}_e$ 
at 6.3~PeV. Re-optimizing the  final energy cut for events interacting through the 
Glashow resonance results in an optimal cut,
$E_{\rm reco} > 0.3$ ~PeV.
No experimental event has been observed in that energy
range, which results in an upper limit on the number of signal events of 
$\mu_{90\%}=2.65$ assuming $\pm$25~\% uncertainties in the signal 
expectation and negligible background expectation.
The limit on the flux at 6.3 PeV is:
\[
\Phi_{\bar{\nu}_e}(E=6.3~{\rm PeV})= 5 \times 10^{-20} {\rm ~GeV^{-1} ~s^{-1} ~sr^{-1} ~cm^{-2}}
.\]

The transformation of this limit to a
limit on a total neutrino flux is not completely straightforward, since
the fraction of ${\overline{\nu}_e}$ produced in the source is unknown.
In cases of neutrino production through  $pp$ collisions one expects a ratio
$\overline{\nu}_e/\nu_e \approx 1$. Hence, one would expect that about 1/6 of
all neutrinos are $\overline{\nu}_e$. However, the $\overline{\nu}_e$ produced 
in 
the $p\gamma \rightarrow n\pi^+$ interaction through decay of the neutron 
carries only a very small 
fraction of the energy, and hence for most neutrino spectra contributes 
negligible to the high energy 
flux of neutrinos. 
 For this case, a flux of $\overline{\nu}_e$ results mainly from neutrino
oscillations. For maximal neutrino mixing,
  $\overline{\nu}_e$ would constitute 1/9 of the total neutrino flux. If
the mixing is non-maximal, that fraction would be smaller.

\section{Conclusion}

We have presented experimental limits on diffuse extragalactic neutrino fluxes.
We find no evidence for neutrino-induced cascades above the backgrounds 
expected from atmospheric neutrinos and muons.  In the energy range  from 
50~TeV to 5~PeV,  the presented limits on the diffuse flux are currently the 
most restrictive.  We have compared our results to several model predictions 
for extragalactic neutrino fluxes and several of these models can be excluded.

Results from the first phase of AMANDA, the 10-string sub-detector AMANDA-B10,
have been reported in \cite{b10} and an update to the analysis 
was presented above. Compared to  AMANDA-B10, the analysis 
presented here has a nearly ten times larger sensitivity, mainly achieved 
through using the larger volume of AMANDA-II and by extending the search to 
neutrinos from all neutrino directions.  

The limits presented here are also more than a factor of two 
below  the AMANDA-B10 limit obtained by searching for neutrino-induced muons
\cite{gary} and roughly as sensitive as the extension of that search using 
AMANDA-II 2000 data \cite{gary_icrc03}. 
(Assuming a neutrino flavor ratio of 1:1:1, the numerical 
limits on the flux of neutrinos of a specific flavor 
(e.q. $\nu_\mu$)
reported in the literature are  1/3 of the limits on the total flux of 
neutrinos.) 
The limits obtained from a search for cascade-like 
events by the Baikal collaboration \cite{baikal} are about 50\% 
less restrictive than the limits presented here. 

With the present analysis one obtains a large sensitivity to astrophysical 
neutrinos of all flavors and in particular to electron and tau neutrinos.  
Hence, given the large sensitivity to muon neutrinos of other  search 
channels, AMANDA can be considered an efficient all-flavor neutrino detector. 

\section{Acknowledgments}

We acknowledge the support of the following agencies: National
Science Foundation--Office of Polar Programs, National Science
Foundation--Physics Division, University of Wisconsin Alumni Research
Foundation, Department of Energy, and National Energy Research
Scientific Computing Center (supported by the Office of Energy
Research of the Department of Energy), UC-Irvine AENEAS Supercomputer
Facility, USA; Swedish Research Council, Swedish Polar Research
Secretariat, and Knut and Alice Wallenberg Foundation, Sweden; German
Ministry for Education and Research, Deutsche Forschungsgemeinschaft
(DFG), Germany; Fund for Scientific Research (FNRS-FWO), Flanders
Institute to encourage scientific and technological research in
industry (IWT), and Belgian Federal Office for Scientific, Technical
and Cultural affairs (OSTC), Belgium;  I.T. acknowledges support from 
Fundaci\'{o}n Venezolana de 
Promoci\'{o}n al Investigador (FVPI), Venezuela;  D.F.C. acknowledges 
the support of the NSF CAREER program.

\end{document}